\documentclass{emulateapj}

\newcommand{\be}{\begin{eqnarray}}
\newcommand{\ee}{\end{eqnarray}}
\newcommand{\lp}{\left(}
\newcommand{\rp}{\right)}
\newcommand{\lb}{\left[}
\newcommand{\rb}{\right]}
\newcommand{\rsh}{r_{\rm sh}}

\begin{document}

\slugcomment{Submitted for publication in The Astrophysical Journal.}
\shorttitle{SUPERNOVA FALLBACK ONTO MAGNETARS}
\shortauthors{PIRO \& OTT}


\title{Supernova Fallback onto Magnetars and Propeller-Powered Supernovae}

\author{Anthony L. Piro and
Christian D. Ott}

\affil{Theoretical Astrophysics, California Institute of Technology, 1200 E California Blvd., M/C 350-17, Pasadena, CA 91125; piro@caltech.edu; cott@tapir.caltech.edu}


\begin{abstract}
We explore fallback accretion onto newly born magnetars during the supernova of massive stars. Strong magnetic fields ($\sim10^{15}\ {\rm G}$) and short spin periods ($\sim1-10\ {\rm ms}$) have an important influence on how the magnetar interacts with the infalling material. At long spin periods, weak magnetic fields, and high accretion rates, sufficient material is accreted to form a black hole, as is commonly found for massive progenitor stars. When $B\lesssim5\times10^{14}\ {\rm G}$, accretion causes the magnetar to spin sufficiently rapidly to deform triaxially and produce gravitational waves, but only for $\approx50-200\ {\rm s}$ until it collapses to a black hole. Conversely, at short spin periods, strong magnetic fields, and low accretion rates, the magnetar is in the ``propeller regime'' and avoids becoming a black hole by expelling incoming material. This process spins down the magnetar, so that gravitational waves are only expected if the initial protoneutron star is spinning rapidly. Even when the magnetar survives, it accretes at least $\approx0.3M_\odot$, so we expect magnetars born within these types of environments to be more massive than the $1.4M_\odot$ typically associated with neutron stars. The propeller mechanism converts the $\sim10^{52}\ {\rm ergs}$ of spin energy in the magnetar into the kinetic energy of an outflow, which shock heats the outgoing supernova ejecta during the first $\sim10-30\ {\rm s}$. For a small $\sim5M_\odot$ hydrogen-poor envelope, this energy creates a brighter, faster evolving supernova with high ejecta velocities  $\sim(1-3)\times10^4\ {\rm km\ s^{-1}}$ and may appear as a broad-lined Type Ib/c supernova. For a large $\gtrsim10M_\odot$ hydrogen-rich envelope, the result is a bright Type IIP supernova with a plateau luminosity of $\gtrsim10^{43}\ {\rm ergs\ s^{-1}}$ lasting for a timescale of $\sim60-80\ {\rm days}$.
\end{abstract}



\section{Introduction}
\label{sec:introduction}

``Magnetars'' are a subset of neutron stars with dipole magnetic fields as strong as $B\sim10^{14}-10^{15}\ {\rm G}$ (Duncan \& Thompson 1992; Thompson \& Duncan 1993). Although at an age of $1,000-10, 000\ {\rm years}$ they have spin periods of $P = 5 -12\ {\rm s}$, as measured from soft gamma-ray repeaters and anomalous X-ray pulsars, it is an outstanding question of how rapidly they rotate when first born. Short initial spin periods ($P_0\sim1-10\ {\rm ms}$) have been favored theoretically so that the dynamo process that creates these strong magnetic fields may operate efficiently \citep{dt92,aki03,tho05}. Motivated by this, many groups have investigated the possible impact of the spindown of this newly formed magnetar in powering an explosion (see, for example, Bodenheimer \& Ostriker 1974; Wheeler et al. 2000; Thompson et al. 2004; Burrows et al. 2007; Dessart et al. 2008). Such short spin periods may also be a source of ultra-high energy cosmic-rays (Arons 2003), create a collimated relativistic flow as needed for gamma-ray bursts \citep[][and references therein]{um07,met10}, or produce a luminous supernova (Kasen \& Bildsten 2010; Woosley 2010).

An assumption of all of these studies is that the supernova which gave birth to the magnetar was successful in ejecting the majority of the progenitor star's envelope. This is clearly correct in many cases, since we know that neutron stars with more modest magnetic fields ($\sim10^{12}\ {\rm G}$) are created in supernovae. But it is possible that some subset of supernovae which produce neutron stars have small injected explosion energies. As is expected for massive stars that give rise to black holes, these would not be successful in ejecting the majority of the envelope and a sizable amount of fallback would occur \citep[as found for $\approx25-40M_\odot$ stars by][]{heg03}. In addition, even in cases where the majority of the envelope is ejected, asymmetries in the explosion may still result in significant fallback. For these reasons, it is plausible that there exists a population of massive stars that give birth to magnetars that are subsequently subject to accretion of the envelope material.

Another motivation for studying fallback accretion onto magnetars is the presence of magnetars near clusters of massive stars. SGR $1806-20$ and CXOU J$164710.2-455216$ are associated with the clusters Cl $1806-20$ and Westerlund 1, respectively, and are inferred to have had progenitor masses of $\approx 40M_\odot$ (Figer et al. 2005; Bibby et al. 2008; Muno et al. 2006). Furthermore, the expanding H I shell around the magnetar 1E $1048.1-5937$ also argues for a $\approx30-40M_\odot$ progenitor (Gaensler et al. 2005). Such massive stars are typically assumed to give rise to black holes \citep{fry99,heg03}, although we note that this will depend sensitively on the details of mass loss during stellar evolution  (Smith et al 2010; O'Connor \& Ott 2011) and on whether these magnetars have binary progenitors (Belczynski \& Taam 2008). It is therefore worth exploring whether the presence of a highly-magnetized neutron star qualitatively changes the outcome of the collapse of massive stars.

In the following study we explore the interaction of newly born magnetars with supernova fallback. We begin in \S \ref{sec:fallback} by discussing the parameter space in which we expect fallback to be important. In \S \ref{sec:spinevolution} we calculate the time-dependent spin evolution of these magnetars. These results are used in \S \ref{sec:magnetarvsbh} to explore whether a newly formed magnetar accretes sufficient material to become a black hole, as a function of the initial spin, magnetic field, and amplitude of the fallback accretion. We also discuss whether these magnetars will be spinning rapidly enough to produce gravitational waves via triaxial instabilities. In \S \ref{sec:propellernova} we show that material expelled in the propeller regime collides with outgoing supernova ejecta, creating a more powerful supernova. We conclude in \S \ref{sec:conclusions} with a summary of our results. In the Appendix we explore the physics of neutrino-cooled accretion columns onto magnetars. 


\section{Fallback versus outflow}
\label{sec:fallback}

Before we investigate the effects of fallback accretion, it is pertinent to discuss when fallback is expected. Although these arguments are strictly applicable for only one-dimension, and we expect a multi-dimensional flow to provide more opportunities for fallback, this gives some intuition about how fallback depends on the accretion rate, spin, and magnetic field strength.

As the rapidly rotating, newly born magnetar spins down, it goes through stages in which it emits energy in dipole spindown radiation and a neutrino-driven, magnetically dominated wind (Thompson, Chang, \& Quataert 2004), both of which may hinder accretion. For a magnetar with a dipole magnetic moment $\mu$ and spin $\Omega$, the spindown luminosity is
\be
	L_{\rm dip} = \frac{\mu^2\Omega^4}{6c^3} = 9.6\times10^{48}\mu_{33}^2P_1^{-4}\ {\rm ergs\ s^{-1}},
	\label{eq:ldip}
\ee
where $\mu_{33}=\mu/10^{33}\ {\rm G\ cm^3}$, as is appropriate for a neutron star with a $10^{15}\ {\rm G}$ magnetic field, and \mbox{$P=2\pi/\Omega=1P_1\ {\rm ms}$}. Assuming this luminosity is carried by a relativistic wind, the associated pressure at a radius $r$ is $p_{\rm dip}=L_{\rm dip}/4\pi c r^2$. Fallback accretion exerts an inward ram pressure, and for the case of spherically symmetric accretion at a rate $\dot{M}$ onto a mass $M$, this is given by
\be
	p_{\rm ram} = \frac{\dot{M}}{8\pi}\lp \frac{2GM}{r^5}\rp^{1/2}.
\ee
Since $p_{\rm dip}\propto r^{-2}$ and $p_{\rm ram}\propto r^{-5/2}$, the spindown luminosity always wins at sufficiently large radii. If the fallback accretion is already proceeding and then the spindown luminosity is to disrupt this accretion flow, we can ask what is the critical accretion rate above which the fallback ram pressure dominates at the magnetar radius $R$. This gives
\be
	\dot{M}_{\rm dip,crit}&=& \frac{\mu^2\Omega^4}{3c^4}\lp \frac{R}{2GM}\rp^{1/2}
	\nonumber
	\\
	&=& 1.8\times10^{-5}\mu_{33}^2P_1^{-4}M_{1.4}^{-1/2}R_{12}^{1/2}\ M_\odot\ {\rm s^{-1}},
	\label{eq:mdotdip}
\ee
where $M_{1.4}=M/1.4M_\odot$ and $R_{12}=R/12\ {\rm km}$. This accretion rate is well-exceeded in all cases we consider.

During the Kelvin-Helmholtz cooling epoch for the newly born magnetar, deleptonization and thermal neutrino losses create a neutrino-driven wind that is magnetically flung by the magnetar's dipole field. For a mass loss rate $\dot{M}_\nu$, the luminosity that goes into this process is (Thompson et al. 2004)
\be
	L_\nu &=& \lp \frac{\mu^2\Omega^4}{\dot{M}_\nu}\rp^{2/5}\dot{M}_\nu
	\nonumber
	\\
	&=& 4.5\times10^{50}\mu^{4/5}P_1^{-8/5}\dot{M}_{\nu,-3}^{3/5}\ {\rm ergs\ s^{-1}},
\ee 
where $\dot{M}_{\nu,-3}=\dot{M}_\nu/10^{-3}M_\odot\ {\rm s^{-1}}$. Repeating the above analysis of assuming this is a relativistic wind and comparing to the ram pressure at the magnetar surface, we derive a critical accretion rate
\be
	\dot{M}_{\nu,\rm crit}&=&\frac{2\dot{M}_\nu}{c}\lp \frac{R}{2GM}\rp^{1/2}\lp \frac{\mu^2\Omega^4}{\dot{M}_\nu}\rp^{2/5}
	\nonumber
	\\
	&=&8.6\times10^{-4}\mu_{33}^{4/5}P_1^{-8/5}\dot{M}_{\nu,-3}^{3/5}M_{1.4}^{-1/2}R_{12}^{1/2}\ M_\odot\ {\rm s^{-1}}.
	\nonumber
	\\
	\label{eq:mdotnu}
\ee
This limit is a little more stringent than the one derived for dipole spindown (eq. [\ref{eq:mdotdip}]). Indeed some of the lower fallback rates we consider are exceeded by this. When Thompson et al. (2004) follow the spindown from a neutrino-driven wind, they find modest amounts of spindown (an increase in the spin period of $\sim5\ {\rm ms}$) even for the most extreme conditions. If there is a phase of spindown from this, it just amounts to different initial conditions from the perspective of our study. Thus, we neglect these effects in our time-dependent spin calculations.

\begin{figure}
\epsscale{1.2}
\plotone{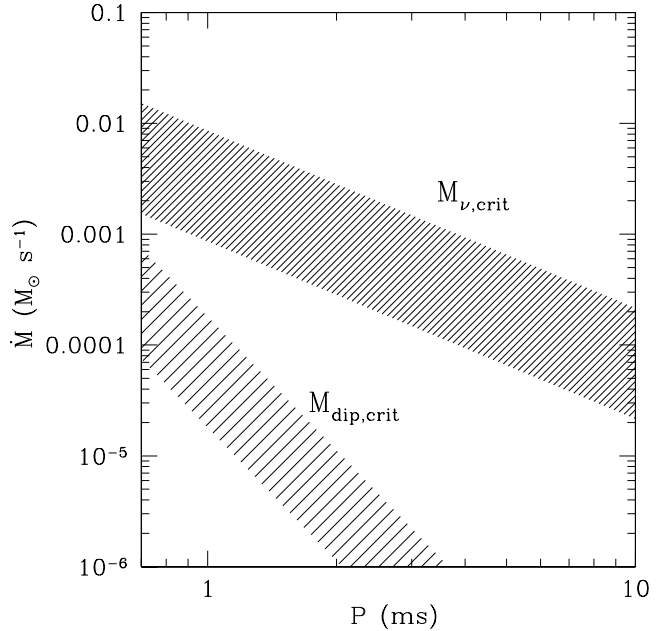}
\caption{The critical accretion rate, above which fallback dominates, as a function of the spin period. We consider two physical processes for inhibiting the fallback: dipole spindown radiation (denoted by $M_{\rm dip,crit}$ and given by eq. [\ref{eq:mdotdip}]), and a neutrino-driven wind (denoted by $M_{\nu,\rm crit}$ and given by eq. [\ref{eq:mdotnu}]). In each case we vary the radius by a factor of 100 (as shown by the shaded regions) to represent uncertainty in the radius at which the accretion flow first comes into contact with this outgoing energy.}
\label{fig:fallbackvsoutflow}
\epsscale{1.0}
\end{figure}

In Figure \ref{fig:fallbackvsoutflow} we summarize the parameter space in which we expect fallback to be important. This shows that the fallback ram pressure dominates for accretion rates above $\sim10^{-5}-10^{-2}M_\odot\ {\rm s^{-1}}$, depending on the process that is inhibiting the fallback. Comparing with the fallback found in numerical studies by MacFadyen et al. (2001) or Zhang et al. (2008), this implies a massive progenitor (in the range of $\sim20-40M_\odot$ for solar metallicity and the progenitor models of Woosley et al. 2002) and a low explosion energy (higher explosion energies lead to weaker fallback; Dessart, Livne, \& Waldman 2010). Although it is not well-known how progenitor mass and explosion energy correlate with magnetar creation, even with these limitations, there is a wide parameter space where fallback onto a magnetar seems inevitable.

Even in cases where these scalings appear to argue that fallback is inhibited, it is still worthwhile to investigate fallback on account that (1) the neutrino-driven wind only lasts $\sim10\ {\rm s}$ while the fallback occurs on a $\gtrsim1000\ {\rm s}$ timescale (reflecting the dynamical time of the progenitor) and (2) the neutrino-driven wind is highly asymmetric. Therefore, even if the wind excavates some region of the progenitor, there is ample opportunity for fallback at other angles. We thus expect that in higher dimensions the strength of fallback is typically greater than what we assume for our one-dimensional arguments.

\section{Spin Evolution Due to Fallback Accretion}
\label{sec:spinevolution}

\subsection{Accretion Versus Expulsion}

The initial spin period of newly-born neutron stars depends on both the spin profile of the progenitor star and subsequent processes that add, subtract, and redistribute angular momentum. Fryer \& Heger (2000) performed smoothed particle hydrodynamics simulations using a rotating progenitor model from Heger et al. (2000) and estimated an initial protoneutron star (PNS) spin period on the order of $100\ {\rm ms}$. It is, however, not clear how they defined the extent of the PNS (see discussion in Ott et al. 2006). The subsequent cooling and contraction to a radius of $\sim12\ {\rm km}$ resulted in $P_0\sim2\ {\rm ms}$. Fryer \& Warren (2004) subsequently estimated neutron star spin periods by assuming that the angular momentum of the inner $1M_\odot$ is conserved as the PNS cools and contracts to a neutron star, finding periods of $\sim1-17\ {\rm  ms}$ depending on the progenitor model. Thompson et al. (2005) studied the action of viscous processes in dissipating the strong rotational shear profile produced by core collapse in a range of progenitors and for different initial iron core periods. They showed that for rapidly rotating cores with postbounce periods of $\lesssim4\ {\rm ms}$, viscosity (presumably due to magnetic torques via the magnetorotational instability or magnetoconvection) spins down the rapidly rotating PNSs by a factor of $\sim2-3$ in the early postbounce epoch. Ott et al. (2006) systematically studied the connection between progenitors and final neutron star spin, generally finding $P_0\sim0.5-10\ {\rm ms}$ and solid body rotation in the PNS core for progenitors with precollapse periods $\lesssim50\ {\rm s}$. We therefore consider initial magnetar spin periods in this range for our present study.

Subsequent to the initial spin period being set as described above, the neutron star may be subject to fallback accretion. Accretion comes under the strong influence of the star's dipole field at the nominal Alfv\'{e}n radius $r_m = \mu^{4/7}(GM)^{-1/7}\dot{M}^{-2/7}$, where $\mu$ is the dipole magnetic moment of the magnetar. For typical magnetar parameters
\be
	r_m = 14\mu_{33}^{4/7}M_{1.4}^{-1/7}\dot{M}_{-2}^{-2/7}\ {\rm km},
	\label{eq:rm}
\ee
where $\dot{M}_{-2}=\dot{M}/10^{-2}M_\odot\ {\rm s^{-1}}$, and the prefactor to $r_m$ can vary depending on the details of the interaction between the flow and magnetic field \citep{gl79,aro86,aro93}. The other critical radius, set by the magnetar's spin $\Omega$, is the corotation radius $r_c=(GM/\Omega^2)^{1/3}$,
\be
	r_c = 17M_{1.4}^{1/3}P_1^{2/3}\ {\rm km}.
\ee
Roughly speaking, one expects that for $r_m<r_c$, material is funneled by the magnetar's dipole field before accreting onto the magnetar's surface, while when $r_m>r_c$, material must spin at a super-Keplerian rate to come into corotation with the magnetar and is thus expelled (the ``propeller regime,'' Illarionov \& Sunyaev 1975). Setting $r_m>r_c$ gives a critical accretion rate
\be
	\dot{M} < 6.0\times10^{-3}\mu_{33}^2M_{1.4}^{-5/3}P_1^{-7/3}\ M_\odot\ {\rm s^{-1}}.
	\label{eq:mdotcrit}
\ee
Comparing to the $25\ M_\odot$ collapsar models of MacFadyen et al. (2001), they find early-time accretion rates of $10^{-4}-10^{-2}\ M_\odot\ {\rm s^{-1}}$ by just varying the injected explosion energy by $(0.255-1.2)\times10^{51}\ {\rm ergs}$. Whether a magnetar is in the propeller regime or not is therefore very sensitive to how energetic the supernova is.

This simplistic picture is not the complete story, as has been detailed by a great many theoretical studies of accretion onto magnetic stars (see for example, Pringle \& Rees 1972; Lynden-Bell \& Pringle 1974; Ghosh \& Lamb 1979; Aly 1980; Wang 1987; Shu et al. 1994; Lovelace et al. 1995, 1999; Ikhsanov 2002; Rappaport et al. 2004; Ek{\c s}i et al. 2005; Kluzniak \& Rappaport 2007; D'Angelo \& Spruit 2010). More recently, numerical simulations have also been used to investigate this problem (Hayashi et al. 1996; Goodson et al. 1997; Miller \& Stone 1997; Fendt \& Elstner 2000; Matt et al. 2002; Romanova  et al. 2003, 2004, 2009). For our present work, we implement a simple model largely based on that used by Ek{\c s}i et al. (2005), as described below. Their prescription has the advantage of being applicable and continuous over a wide range of parameters, while capturing the main expected features of the propeller regime.

In cases where $r_c>r_m>R$, the inflowing material is channeled onto the magnetar poles where it shocks and neutrino cools. We save a more detailed treatment of the physics of this process for the Appendix, since it does not have a direct bearing on our results for the time-dependent spin, which we consider next.

\subsection{Time-Dependent Spin From Fallback Accretion}

Given this picture of accretion and expulsion described above, we solve for spin evolution under the influence of fallback accretion by integrating the differential equation
\be
	I\frac{d\Omega}{dt} = N_{\rm dip} + N_{\rm acc},
	\label{eq:spin_differential}
\ee
where $I=0.35MR^2$ is the moment of inertia (Lattimer \& Prakash 2001), and $N_{\rm dip}$ and $N_{\rm acc}$ are the torques from dipole emission and accretion, respectively. As discussed in \S \ref{sec:fallback}, we ignore spindown from neutrino-driven winds in equation (\ref{eq:spin_differential}). The dipole spindown torque is given by
\be
	N_{\rm dip} = -\frac{\mu^2\Omega^3}{6c^3} = -1.5\times10^{45}\mu_{33}^2P_1^{-3}\ {\rm ergs}.
\ee
We assume that the magnetar is rotating as a solid body, as is likely the case within $\sim1\ {\rm s}$ of collapse since the MRI \citep{tho05,ott06} or low-$T/|W|$ instabilities \citep{wat05,ott05} will limit differential rotation. When $r_m>R$, material leaves the disk with the specific angular momentum at a radius $r_m$. Depending on the relative positions of the Alfv\'{e}n and corotation radii, this can either spin up or spin down the magnetar, so we write the torque as
\be
	N_{\rm acc} = 
			n(\omega)(GMr_m)^{1/2}\dot{M} \quad {\rm  if}\ r_m>R,
			\label{eq:nacc}
\ee
where $n(\omega)$ is the dimensionless torque which depends on the fastness parameter $\omega = \Omega/(GM/r_m^3)^{1/2} = (r_m/r_c)^{3/2}$. Ek{\c s}i et al. (2005) discuss different ways in which $n(\omega)$ can be set, but for simplicity we take $n=1-\omega$. This has the advantage that the torque goes to zero at the corotation radius, is continuous for all $\omega$, and goes negative when $r_m>r_c$, corresponding to the spin down which occurs during the propeller regime. As $\omega$ gets larger, this prescription gives increasingly strong spindown, consistent with the more detailed simulations of Romanova et al. (2004). When $r_m<R$ we set the torque to
\be
	N_{\rm acc} = 
			\left( 1- \Omega/\Omega_{\rm K}\right)(GMR)^{1/2}\dot{M}  \quad  {\rm  if}\ r_m<R,
\ee
where $\Omega_K=(GM/R^3)^{1/2}$. The prefactor is included to ensure that torque is continuous for all values of $r_m$. The disadvantage is that since the prefactor is $\lesssim1$, it will underpredict the amount of torque, but this does not change our main conclusions, as we discuss in \S 4.2.

As we integrate the spin in time, we keep track of the magnetar's rotation parameter, $\beta\equiv T/|W|$, where $T=I\Omega^2/2$.  We use the prescription given in Lattimer \& Prakash (2001) for $|W|$,
\be
	|W|\approx 0.6Mc^2\frac{GM/Rc^2}{1-0.5(GM/Rc^2)}.
\ee
We keep $R$ fixed even as $M$ changes, which is roughly consistent with most equations of state, except when $M$ gets near its maximum value (Lattimer \& Prakash 2001). When $\beta=0.5$, the neutron star is at breakup and cannot accept further angular momentum. Even prior to this, dynamical bar-mode instabilities occur for $\beta>0.27$ (Chandrasekhar 1969), and secular instabilities for $\beta\gtrsim0.14$, driven by gravitational radiation reaction or viscosity (Lai \& Shapiro 1995). Since the dynamical bar-mode instability is guaranteed to radiate and/or hydrodynamically re-adjust angular momentum, we set $N_{\rm acc}=0$ when $\beta>0.27$. We ignore changes in spin due to the secular instabilities since growth timescales are uncertain and may be suppressed by competition between viscosity or gravitational radiation reaction \citep{ls95}.

We parameterize the fallback accretion rate to mimic the results of MacFadyen et al. (2001) and Zhang et al. (2008). This can roughly be broken into two parts. At early times it scales as
\be
	\dot{M}_{\rm early} = \eta 10^{-3}t^{1/2}M_\odot\ {\rm s^{-1}},
	\label{eq:mdotearly}
\ee
where $\eta\approx0.1-10$ is a factor that accounts for different explosion energies (a smaller $\eta$ corresponds to a larger explosion energy), and $t$ is measured in seconds. The late time accretion is roughly independent of the explosion energy and is set to be
\be
	\dot{M}_{\rm late} = 50t^{-5/3}M_\odot\ {\rm s^{-1}}.
	\label{eq:mdotlate}
\ee
The accretion rate at any given time is found from combining these two expressions
\be
	\dot{M} = \lp \dot{M}_{\rm early}^{-1} + \dot{M}_{\rm late}^{-1}  \rp^{-1}.
	\label{eq:mdot}
\ee
The mass of the neutron star increases at a rate $\dot{M}$ when $r_m<r_c$ and is set fixed when $r_m>r_c$. For comparison, we also integrate $\dot{M}$ for all values of $r_m$ to follow how much matter the magnetar would have accreted if not for the propeller mechanism.

Equation (\ref{eq:mdot}) reflects fallback of the envelope, but most likely this material must pass through a disk before finally accreting onto the magnetar. To test this hypothesis and explore whether this leads to a quantitative change of the accretion rate, we built one-zone, $\alpha$-disk models (similar to Metzger et al. 2008) using the angular momentum profiles of the massive, rotating progenitors of Woosley \& Heger (2006) simulated with GR1D (O'Connor \& Ott 2010). Our general finding was that (1) there is sufficient angular momentum to form a disk, and
(2) the disk is nearly steady-state, where the accretion rate onto the star differs from the infall rate by no more than a factor of $\sim5$ (and this scales with the $\alpha$-viscosity, with a larger $\alpha$ resulting in higher accretion rates), and (3) the radius of the disk is typically well outside of the Alfv\'{e}n radius. We therefore consider the mediation of the disk to be degenerate with $\eta$ and use the direct infall rates as described above.

\begin{figure}
\epsscale{1.2}
\plotone{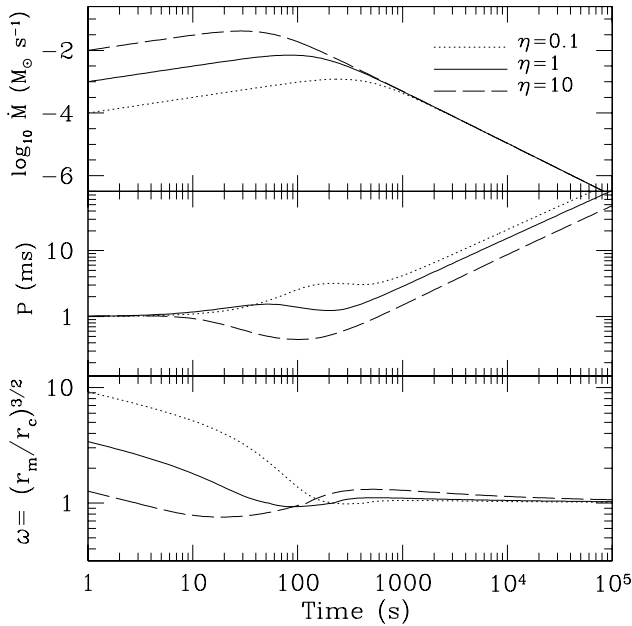}
\caption{The spin evolution of a magnetar with $B=10^{15}\ {\rm G}$ and an initial spin period of $P_0=1\ {\rm ms}$. We compare values of $\eta=0.1,1$, and $10$, demonstrating the strong effect early-time accretion can have. The top panel shows the time-dependent accretion rate, the center panel shows the spin period, and the bottom panel shows the fastness parameter $\omega$, where $\omega>1$ corresponds to the propeller regime and $\omega\leq1$ corresponds to accretion.}
\label{fig:evolution}
\epsscale{1.0}
\end{figure}

In Figure \ref{fig:evolution}, we compare integrations of equation (\ref{eq:spin_differential}) for values of $\eta=0.1,1,$ and $10$. The top panel shows the accretion rate given by equation (\ref{eq:mdot}). The middle panel plots the time-dependent spin period. The bottom panel plots the fastness parameter, which reflects whether or not the magnetar is in the propeller regime. For $\eta=0.1$, only $0.25\ M_\odot$ is accreted out of a potential amount of accretion of $1.55\ M_\odot$, and for $\eta=1$ only $1.03\ M_\odot$ is accreted out of a potential amount of $3.15\ M_\odot$. Therefore both these cases are able to avoid becoming a black hole via the propeller mechanism (assuming a maximum neutron star mass of $2.5M_\odot$). In contrast, the $\eta=10$ case (which corresponds to a lower-energy explosion) accretes $3.45\ M_\odot$ out of $6.41\ M_\odot$, which means it likely becomes a black hole. Since the accretion rate is highest at early times, black hole formation happens rather quickly during the runs, at $\approx34\ {\rm s}$ and $\approx46\ {\rm s}$ for maximum neutron star masses of $2.5M_\odot$ and $3M_\odot$, respectively.

In each of these cases, the spin eventually reaches an equilibrium value that simply tracks $\dot{M}$ with $\omega\approx1$. Setting $r_m=r_c$, we calculate an equilibrium spin period,
\be
	P_{\rm eq} &=& 2\pi\mu^{6/7}(GM)^{-5/7}\dot{M}^{-3/7}
	\nonumber
	\\
	&=& 5.8\mu_{33}^{6/7}M_{1.4}^{-5/7}\dot{M}_{-4}^{-3/7}\ {\rm ms},
	\label{eq:peq}
\ee
where $\dot{M}_{-4}=\dot{M}/10^{-4}M_\odot\ {\rm s^{-1}}$.


\section{Magnetar versus black hole formation}
\label{sec:magnetarvsbh}

\subsection{The Amount of Mass Accreted}

\begin{figure}
\epsscale{1.2}
\plotone{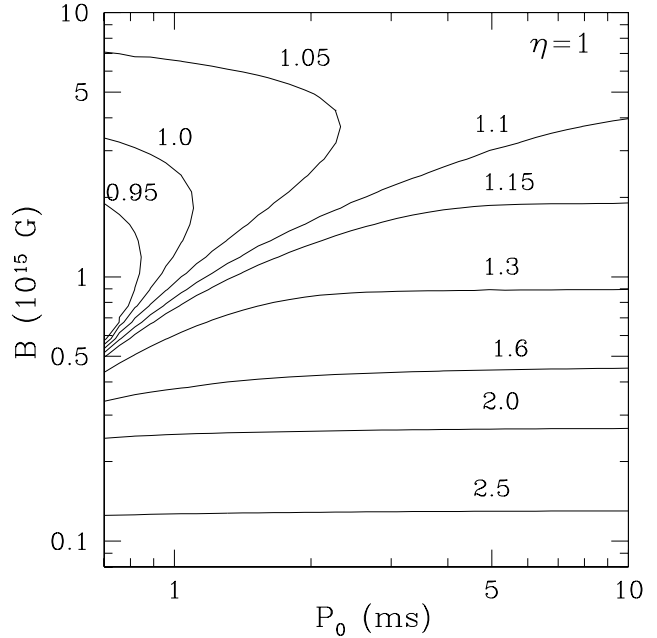}
\caption{Contours show the amount of mass accreted (in solar masses) for different initial spin periods and magnetic field strengths (all for $\eta=1$). For every case we assume an initial magnetar mass of $1.4M_\odot$ with a radius of $12\ {\rm km}$. If the propeller regime did not expel material, then $3.15M_\odot$ would have been accreted.}
\label{fig:contour}
\epsscale{1.0}
\end{figure}

\begin{figure}
\epsscale{1.2}
\plotone{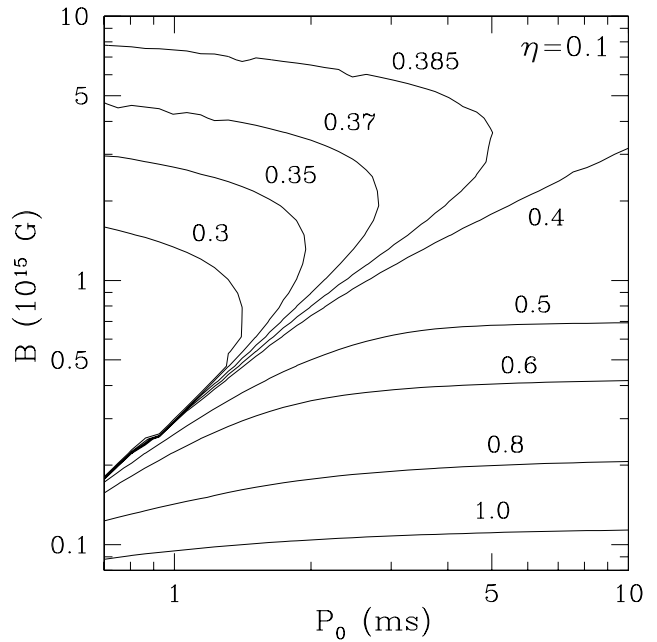}
\caption{The same as Fig. \ref{fig:contour}, but with $\eta=0.1$. For this case, if the propeller regime did not expel material, then $1.55M_\odot$ would have been accreted.}
\label{fig:contour2}
\epsscale{1.0}
\end{figure}

The example models in the previous section demonstrate that the amount of mass accreted by the magnetar depends strongly on whether the propeller regime is reached. Therefore, whether or not a magnetar eventually becomes a black hole depends on its initial spin period and magnetic field. This is in stark contrast to neutron stars with dynamically unimportant magnetic fields whose fates simply depend on the properties of the supernova and the compactness of the stellar core \citep{zha08,oo11}. To explore these correlations, we plot contours for the amount of mass accreted as a function of the initial spin period and magnetic field in Figures \ref{fig:contour} and  \ref{fig:contour2} for values of $\eta=1$ and $0.1$, respectively. In the $\eta=1$ case, a magnetar remains for only a small fraction of the initial conditions (this of course depends on the value of the maximum neutron star mass). For $\eta=0.1$, a magnetar is expected for the majority of the parameter space. Since these two values of $\eta$ correspond to a factor of $\sim2$ difference in the initial explosion energy \citep{mac01}, these comparisons demonstrate just how sensitive the outcome is to this quantity.

\begin{figure}
\epsscale{1.2}
\plotone{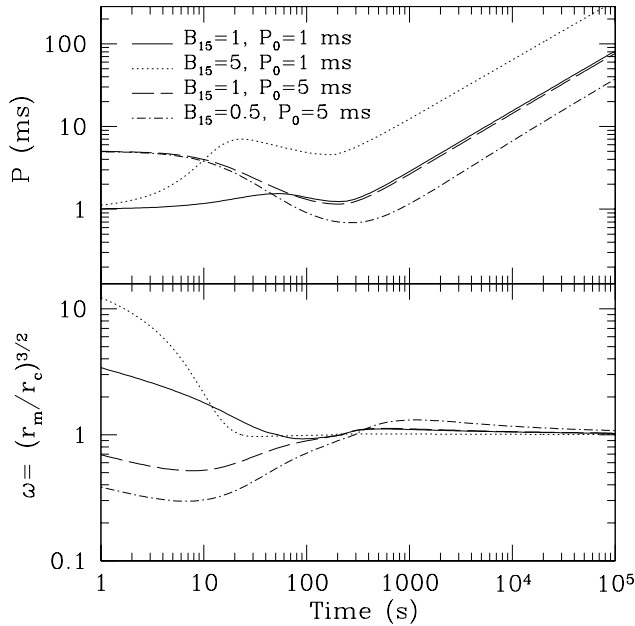}
\caption{Time-evolution of the spin period and fastness parameter $\omega$ for a diverse selection of models. The accretion rate corresponds to $\eta=1$ for all cases (the solid line from the top panel in Fig. \ref{fig:evolution}). The low magnetic field ({\it dot-dashed lines}) and slowly spinning ({\it dashed line}) cases exceed a mass of $2.5M_\odot$ at $\approx180\ {\rm s}$, at which point they most likely become black holes. See the text for further discussion of the features exhibited here.}
\label{fig:examples}
\epsscale{1.0}
\end{figure}

The general trend is that at high magnetic fields and small periods, there is less mass accretion due to the propeller mechanism being stronger. Nevertheless, there are also some subtle differences from this trend that are due to the interaction of a time-dependent accretion rate with the changing spin. For example, in Figure \ref{fig:contour} we see that the minimum accreted mass occurs near $B\approx10^{15}\ {\rm G}$ and $P_0\approx0.7\ {\rm ms}$, and the accreted mass actually {\it increases} for stronger magnetic fields, contrary to our intuition for when the propeller mechanism  should be strongest. To explore what is happening here, we plot the spin evolution for a collection of different magnetic fields and initial spin parameters in Figure \ref{fig:examples}.  We can see that at sufficiently strong magnetic fields, the propeller is so strong that the star quickly spins down during the first $\approx20\ {\rm s}$ ({\it dotted line}). At this point the accretion rate has increased dramatically, and the star now accretes and spins up until about $\approx200\ {\rm s}$. It is due to this stage that the magnetar accretes more than was expected.

One takeaway message of this parameter survey is that for all these models at least $\approx0.3M_\odot$ is accreted. Therefore magnetars that are subject to the conditions of being born within a massive star should on average be more massive than the $1.4M_\odot$ typically associated with neutron stars. Measuring the masses of magnetars would therefore be useful for constraining whether some are indeed born in massive progenitors.

\subsection{Prospects for Gravitational Wave Production}

\begin{figure}
\epsscale{1.2}
\plotone{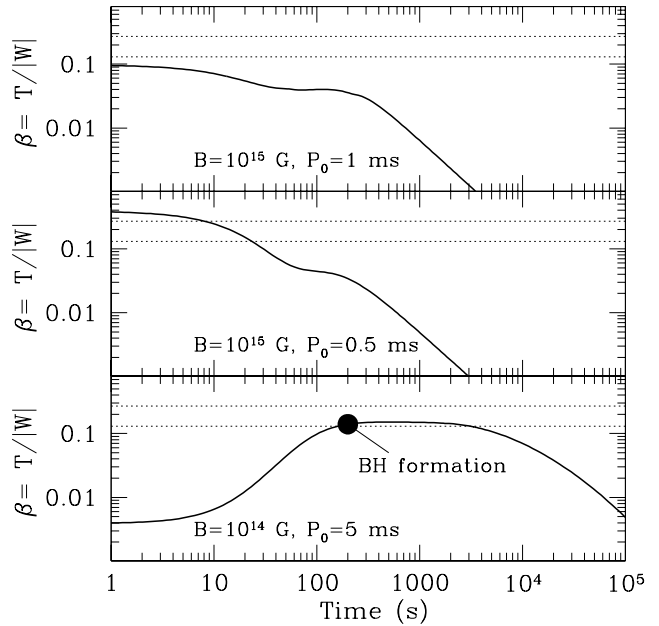}
\caption{The spin parameter $\beta$ for a selection of models, all with $\eta=1$. The dotted lines denote $\beta=0.14$ (the critical value for secular instabilities) and $\beta=0.26$ (the critical value for dynamical instabilities). Gravitational wave emission is only expected above the dotted lines, and is generally seen for extremely short initial spin periods ($P_0\lesssim0.7\ {\rm ms}$) or small magnetic field strength($B\lesssim5\times10^{14}\ {\rm G}$).}
\label{fig:spin}
\epsscale{1.0}
\end{figure}

These results also have bearing on whether a young magnetar should be expected to be an important gravitational wave source (as discussed in Corsi \& M{\'e}sz{\'a}ros 2009, and references therein). For this to occur, it must be spinning sufficiently quickly that dynamical bar-mode instabilities or secular instabilities are excited. To explore this, we plot the spin parameter $\beta$ for a selection of models in Figure \ref{fig:spin}. The majority of the parameter space we probe experiences some time in the propeller regime, spinning down the magnetar, and making gravitational wave emission unlikely. For magnetic fields $\lesssim5\times10^{14}\ {\rm G}$, the magnetar is spun up by accretion sufficiently to produce gravitational waves, but the accretion then quickly leads to collapse to a black hole. This is seen in the bottom panel of Figure \ref{fig:spin}, where $\beta>0.14$ for a time, but  then exceeds a mass of $2.5M_\odot$ at $\approx180\ {\rm s}$. This model never exceeds $\beta=0.27$, but this is an artificial effect of the $1-\Omega/\Omega_{\rm K}$ factor for the torque prescription (see eq. [10]). If we instead assume the magnetar accreted with the specific angular momentum at its surface of $(GMR)^{1/2}$, $\beta=0.27$ would be easily reached. We estimate the timescale for gravitational wave emission by integrating the early time accretion law,
\be
	t_{\rm gw} = 140\eta^{-2/3}\lp\frac{M_{\rm max}-M_0}{1.1M_\odot} \rp^{2/3}{\rm s},
	\label{eq:tgw}
\ee
where $M_{\rm max}$ is the maximum neutron star mass before black hole formation and $M_0$ is the initial neutron star mass. The accretion peaks on a timescale
\be
	t_p = 150\eta^{-6/13}\ {\rm s},
	\label{eq:tp}	
\ee
which is found by equating equations (\ref{eq:mdotearly}) and (\ref{eq:mdotlate}). So our two conditions for equation (\ref{eq:tgw}) to be valid are that $B\lesssim5\times10^{14}\ {\rm G}$ and $t_{\rm gw}<t_p$. If $B\gtrsim5\times10^{14}\ {\rm G}$ we don't expect appreciable spinup and gravitational wave emission, and if $t_{\rm gw}>t_p$ then the gravitational wave emission timescale is merely $\approx t_p$.


\section{Propeller-powered supernovae}
\label{sec:propellernova}

In cases that do not collapse into black holes, the material expelled by the propeller mechanism collides with the supernova ejecta. This shock-heats the envelope and increases the energy budget of the supernova. We next estimate the observable signature of such powering. The process we describe here is decidedly different from what was explored by Kasen \& Bildsten (2010) and Woosley (2010), who used dipole spindown luminosity to heat and power a more luminous supernova. In their case the dipole spindown takes place on sufficiently long timescales that it can directly power an extremely luminous supernova. As we discuss below, the majority of the energy from the propeller-mechanism is injected during the first $\sim10-30\ {\rm s}$, so it can be treated as a sudden impulse of energy at early times. The majority of this energy is therefore lost to adiabatic expansion and not seen directly in the peak luminosity. Nevertheless, the energy can accelerate the supernova ejecta to high velocities of up to $\sim(1-3)\times10^4\ {\rm km\ s^{-1}}$, which are observable in the spectra and alter the lightcurve shape.

\subsection{Propeller Energy Budget}

The expelled material carries a kinetic energy equal to the spindown energy of the magnetar. To help power the supernova, this material must climb out of the magnetar's gravitational well, so we estimate the kinetic luminosity of the propeller material as
\be
	L_{\rm prop}
	= -N_{\rm acc}\Omega- GM\dot{M}/r_m,
	\label{eq:lprop}
\ee
where the negative sign in the first term is because we have defined $N_{\rm acc}$ to be negative when the magnetar is spinning down (eq. [\ref{eq:nacc}]). With this equation we have assumed that the majority of the outflow originates from the inner edge of the disk. While this is a reasonable assumption, it also means that the material has to travel the furthest out of the potential well. If material can leave the disk at larger radii, it will require less energy to do so, thus this represents a lower limit. The total energy that can possibly be put into expelled material is limited by the magnetar rotation,
\be
	E_{\rm rot} =\frac{1}{2}I\Omega^2= 2.8\times10^{52}M_{1.4}R_{12}^2P_1^{-2}.
\ee
In some cases the early-time accretion may even spin the magnetar up to sub-millisecond spin periods before the propeller mechanism begins. In these cases the magnetar stores the accretion energy in its spin, which is tapped via the propeller mechanism to help power the supernova.

\begin{figure}
\epsscale{1.2}
\plotone{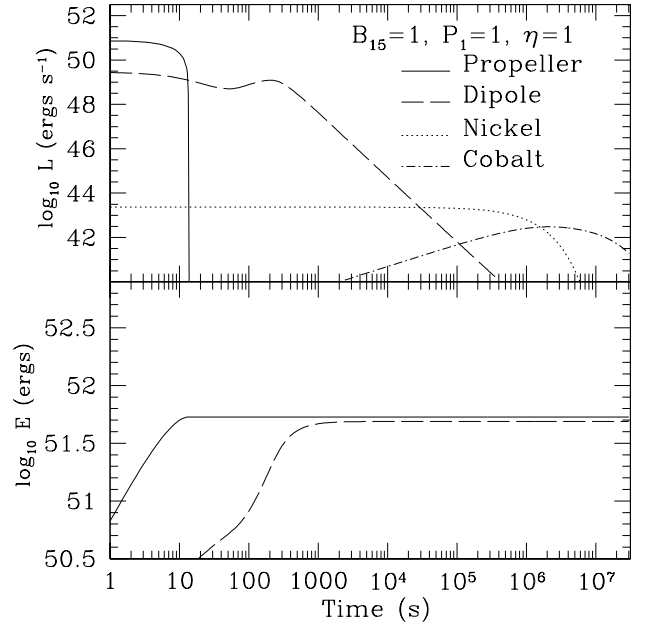}
\caption{The luminosity emitted by the magnetar, either from kinetic energy of magnetically expelled material $L_{\rm prop}$ ({\it solid lines}), dipole radiation $L_{\rm dip}$ ({\it dashed lines}), and, for comparison, decay of $0.3M_\odot$ of $^{56}$Ni ({\it dotted lines}). The initial parameters are $P_0=1\ {\rm ms}$ with $\eta=1$ with a magnetic field of $10^{15}\ {\rm G}$. The bottom panel shows the integrated energy as a function of time for each case. The total energy from radioactive heating is not sufficient to appear on the bottom panel.}
\label{fig:energy_injection}
\epsscale{1.0}
\end{figure}

\begin{figure}
\epsscale{1.2}
\plotone{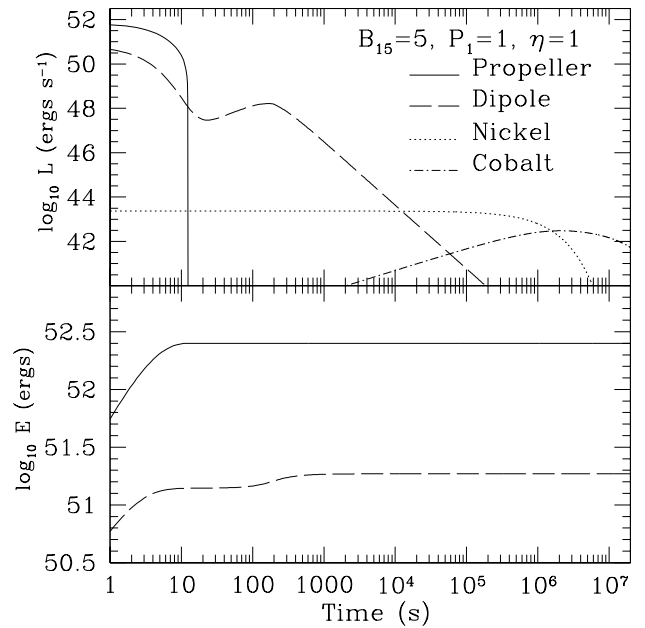}
\caption{The same as Fig. \ref{fig:energy_injection}, but with a magnetic field of $5\times10^{15}\ {\rm G}$. This example is more strongly in the propeller-regime at early times, so that a larger fraction of the spin energy goes into the kinetic energy of the outflow.}
\label{fig:energy_injection2}
\epsscale{1.0}
\end{figure}

In Figures \ref{fig:energy_injection} and \ref{fig:energy_injection2} we quantify the luminosity of the propeller mechanism as well as what fraction of $E_{\rm rot}$ is able to be tapped by this process. The top panels of each figure show $L_{\rm prop}$ (eq. [\ref{eq:lprop}]), $L_{\rm dip}$ (eq. [\ref{eq:ldip}]), and the radioactive decay of $0.3M_\odot$ of $^{56}$Ni as a function of time. The propeller-powering only lasts $\sim10-30\ {\rm s}$ until $r_m\sim r_c$. At this point the magnetar is not spinning sufficiently rapidly to expel material to infinity and the luminosity quickly shuts off. The bottom panels of Figures \ref{fig:energy_injection} and \ref{fig:energy_injection2} show the integrated energy as a function of time,
\be
	E_i(t) = \int_0^t L_i(t) dt,
\ee
where $i$ stands for either the propeller luminosity or dipole luminosity. In Figure \ref{fig:energy_injection}, less than $\sim20\%$ of the rotational energy goes into expelling material. For a stronger magnetic field the propeller regime is more extreme, and nearly all of the rotational energy is converted into energy of outflowing material, as shown in Figure \ref{fig:energy_injection2}. In either case, this additional energy may be greater than the typical supernova energy $E_{\rm sn}$ of $\sim10^{51}\ {\rm ergs}$. In the following sections we explore how this additional energy alters the properties of the supernova depending on the mass of the envelope material, representative of Type Ib/c and Type IIP supernovae.

\subsection{Low-Mass Envelopes}

We consider a supernova with initial energy $E_{\rm sn}$, ejecta mass $M_{\rm ej}$, subject to a sudden impulse of energy $E_{\rm prop}$. As we will show, the observational impact of $E_{\rm prop}$ depends strongly on the ejecta mass and composition. In this section we focus on the properties of an event with a hydrogen-deficient envelope and a mass $M_{\rm ej}\lesssim5M_\odot$, as is expected for the progenitors of Type Ib/c supernovae that have lost a large fraction of their envelope (including all of their hydrogen) to stellar winds, binary mass transfer, and/or outbursts \citep{smi11}.

The collision of the propeller material with the supernova ejecta shock heats and accelerates the ejecta. For a total energy $E_{\rm tot}=E_{\rm sn}+E_{\rm prop}$, the final velocity, with which it coasts for the remainder of the expansion, is
\be
	v_f \approx (2E_{\rm tot}/M_{\rm ej})^{1/2}
	= 2500E_{52.5}^{1/2}M_{5}^{-1/2}\ {\rm km\ s^{-1}},
	\nonumber
	\\
	\label{eq:vf}
\ee
where $E_{52.5}=E_{\rm tot}/3\times10^{52}\ {\rm ergs}$ and $M_5=M_{\rm ej}/5M_\odot$. The diffusion timescale of photons from this hot, expanding material is given by
\be
	t_d= \lp \frac{M_{\rm ej}\kappa}{13.78v_fc} \rp^{1/2}
	= 11\kappa_{0.1}^{1/2}M_{5}^{3/4}E_{52.5}^{-1/4}\ {\rm days},
	\label{eq:td}
\ee
where $\kappa$ is the opacity, which we scale to $\kappa_{0.1}=\kappa/0.1\ {\rm cm^2\ g^{-1}}$ (the typical opacity used for a gray calculation, Pinto \& Eastman 2000), and the factor of 13.78 comes from detailed analytic studies of Type I supernovae \citep{arn82,pe00}. The shell becomes optically thin on a timescale
\be
	t_\tau\approx \lp \frac{3}{4\pi}\frac{M_{\rm ej}\kappa}{v_f^2} \rp^{1/2}
	=  226\kappa_{0.1}^{1/2}M_5E_{52.5}^{-1/2}\ {\rm days}.
\ee
The diffusion approximation we will use is not applicable after this time. The increased velocity creates a faster and more luminous supernova, but this higher luminosity is not directly from energy input from the propeller mechanism. Instead, since the explosion velocity is higher, the diffusion time (eq. [\ref{eq:td}]) is shorter, and the $^{56}$Ni decay is being probed at earlier times.

To understand the corresponding lightcurve created by propeller energy being injected into the explosion, we construct a simple, one-zone model of the expansion, cooling and emission, following the mathematical framework of Li \& Paczy\'{n}ski (1998; also see Kulkarni 2005, Kasen \& Bildsten 2010). For an expanding shell, the internal energy $E_{\rm int}$ satisfies the differential equation (rewritten from eq. [9] of Li \& Paczy\'{n}ski 1998)
\be
	\frac{1}{t}\frac{d}{dt}\lb E_{\rm int}(t)t\rb = L_{\rm prop}(t)+L_{\rm nuc}e^{-t/t_\tau}-L(t),
	\label{eq:diffusion}
\ee
where
\be
	L(t)=E_{\rm int}(t)t/t_d^2
	\label{eq:luminosity}
\ee
is the emitted luminosity. The nuclear luminosity includes contributions from $^{56}$Ni decay and subsequent $^{56}$Co decay.
We use the analytic expression \citep{pe00,bbh11}
\be
	L_{\rm nuc}(t) &=& \epsilon_{\rm Ni} M_{\rm Ni} e^{-t/t_{\rm Ni} }
	+ \epsilon_{\rm co} M_{\rm Ni}\left[ e^{-t/t_{\rm Co}}  - e^{-t/t_{\rm Ni}} \right],
	\nonumber
	\\
	\label{eq:lni}
\ee
where $M_{\rm Ni}$ is the mass of $^{56}$Ni synthesized, $\epsilon_{\rm Ni}=3.9\times10^{10}\ {\rm ergs\ g^{-1}\ s^{-1}}$,  $t_{\rm Ni}= 7.6\times10^{5}\ {\rm s}$,  $\epsilon_{\rm Co}=6.8\times10^{9}\ {\rm ergs\ g^{-1}\ s^{-1}}$, and $t_{\rm Co}=9.8\times10^6\ {\rm s}$. The factor of $e^{-t/t_\tau}$ in equation (\ref{eq:diffusion}) takes into account that the material eventually becomes optically thin to gamma-rays (although this factor only leads to small changes at late times). We have tested this simplified model against a wide range of nuclear-powered explosion calculations \citep{ew88,imn98,dar10}, and found qualitatively good fits to the timescales and magnitudes of the peak luminosity.

\begin{figure}
\epsscale{1.2}
\plotone{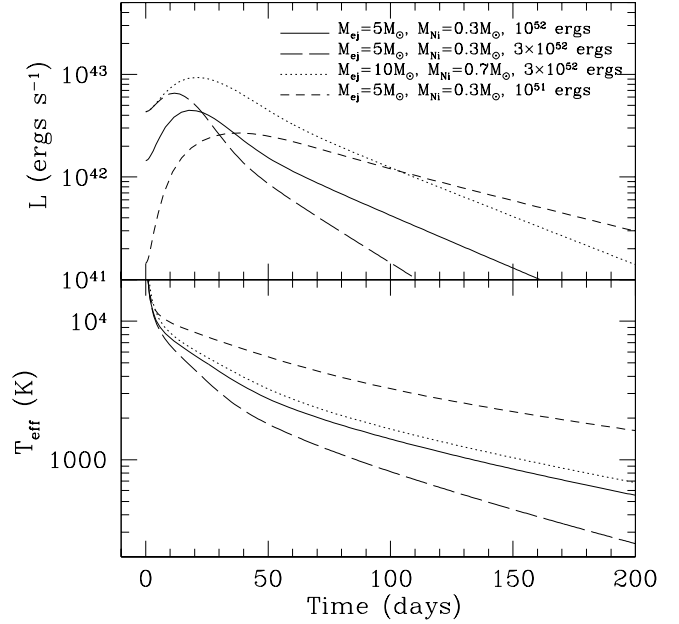}
\caption{Bolometric luminosities and effective temperatures calculated using the one-zone model described by eq. (\ref{eq:diffusion}). Each curve is labeled with a $M_{\rm ej}$, $M_{\rm Ni}$, and the total energy input. See text for further details. The general trend is that the energy injection of the propeller mechanism creates a brighter, more quickly evolving supernova.}
\label{fig:supernova2}
\epsscale{1.0}
\end{figure}
For Figure \ref{fig:supernova2}, we integrate equation (\ref{eq:diffusion}) numerically, setting $E_{\rm int}=E_{\rm sn}=10^{51}\ {\rm ergs}$ at $t=0$. Each curve is labeled by different values for $M_{\rm ej}$, $M_{\rm Ni}$, and the total energy input (initial supernova energy plus the propeller mechanism).  The initial radius is $5R_\odot$ as appropriate for a compact Wolf-Rayet progenitor. We set $T_{\rm eff}=(L/4\pi r^2\sigma)^{1/4}$, where $r=v_ft$ and $\sigma$ is the Stefan-Boltzmann constant. This temperature is only accurate up to a time $t\approx t_\tau$. The first two models ({\it solid and long-dashed lines}) explore the effect of a high input energy. The curve labeled with $M_{\rm ej}=10M_\odot$ and $M_{\rm Ni}=0.7M_\odot$ ({\it dotted line}) is representative of a ``hypernova'' model \citep{nom01}. For this model we use an initial radius of $10R_\odot$, consistent with massive helium stars \citep{woo95}. The model with merely $10^{51}\ {\rm ergs}$ ({\it short-dashed line}) is meant to be representative of a normal Type Ib/c supernova.

From these calculations, we find the general trend that additional energy injection from the propeller mechanism creates a brighter, more quickly evolving supernova (it will also cool faster and show optically thin features sooner). Within the framework we have described, it is not necessary that these events produce more $^{56}$Ni than average. We therefore expect propeller-powered Type Ib/c supernovae to be associated with a wide range of peak luminosities, but to generically exhibit high velocities of $\sim(1-3)\times10^4\ {\rm km\ s^{-1}}$.

\subsection{High-Mass Envelopes}

If the envelope is more massive and has a hydrogen-rich composition, the lightcurve evolution can be significantly different, as is seen for Type IIP supernovae. The analytic features of these lightcurves were well summarized by \citet{pop93}, whose work was confirmed and expanded upon by the numerical simulations of \citet{eas94} and \citet{kw09}, and also studied with the first non-LTE time-dependent radiative-transfer simulations by \citet{des11}. The general picture is that the backward progression of a hydrogen recombination wave through the expanding ejecta causes the supernova to radiate at a fixed effective temperature set by the ionization temperature $T_{\rm eff}=2^{1/4}T_{\rm ion}$. This continues until the entire envelope has become neutral, which truncates the luminosity, revealing $^{56}$Co decay if it is sufficiently available. \citet{pop93} demonstrated that a hydrogen-rich envelope will exhibit a plateau phase when a certain dimensionless parameter is greater than unity. We rewrite this condition in terms of a critical mass, finding
\be
	M_{\rm ej} \gtrsim 6E_{52.5}^{1/3}R_{500}^{2/3}\kappa_{0.34}^{-4/3}T_{5045}^{-8/3}M_\odot,
\ee
where $\kappa_{0.34}=\kappa/0.34\ {\rm cm^2\ g^{-1}}$ is the electron-scattering opacity for a solar composition, $R_0$ is the initial stellar radius with $R_{500}=R_0/500R_\odot$, and $T_{5045}=T_{\rm ion}/5045\ {\rm K}$, corresponding to an effective temperature of $6000\ {\rm K}$. For ejecta masses larger than this we expect a prominent plateau phase and, scaling the analytic results of \citep{pop93} to our values, with a plateau luminosity of
\be
	L_{\rm plat} = 2.8\times10^{43}M_{10}^{-1/2}E_{52.5}^{5/6}R_{500}^{2/3}\kappa_{0.34}^{1/3}T_{5045}^{4/3}\ {\rm ergs\ s^{-1}},
	\nonumber
	\\
	\label{eq:lplat}
\ee
and a plateau timescale
\be
	t_{\rm plat} = 56M_{10}^{1/2}E_{52.5}^{-1/6}R_{500}^{1/6}\kappa_{0.34}^{1/6}T_{5045}^{-2/3}\ {\rm days},
	\label{eq:tplat}
\ee
where we note that \citet{kw09} find a slightly stronger scaling of $t_{\rm plat}\propto E^{-1/4}$ in their numerical results. The large energy input would also result in higher velocities of $v_f\sim10^4\ {\rm km\ s^{-1}}$ (scaling eq. [\ref{eq:vf}] to a mass of $\sim10-20M_\odot$), which although not as high as in broad-lined supernovae, would be anomalously high for a Type IIP supernova. The high luminosities we find are similar to what is seen for many Type IIn supernovae (see Fig. 3 of Smith et al. 2008), but our events would {\it not} have nebular features from the interaction with winds and thus would appear distinct from Type IIn supernovae.

\begin{figure}
\epsscale{1.2}
\plotone{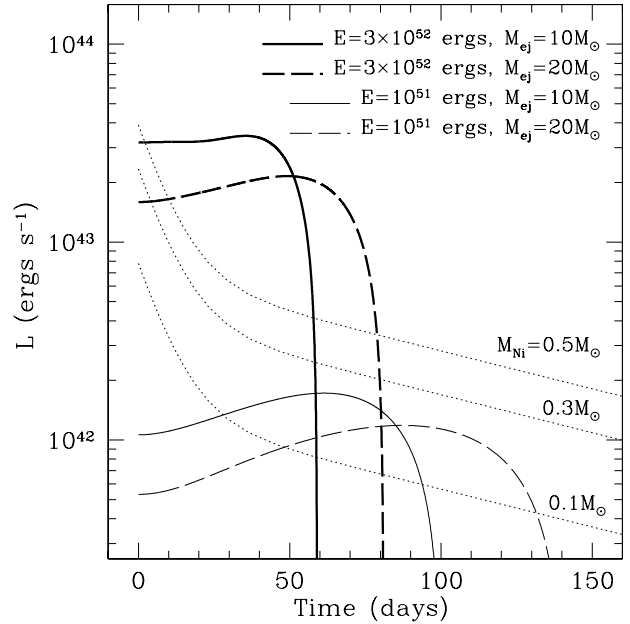}
\caption{Bolometric luminosity calculated using the analytic model of \citet{pop93}. The thick lines are models with an energy injection of $3\times10^{52}\ {\rm ergs}$, and the thin lines use $10^{51}\ {\rm ergs}$ (representative of a normal Type IIP supernova). The additional energy injection results in a larger luminosity, but shorter timescale, consistent with eqs. (\ref{eq:lplat}) and (\ref{eq:tplat}). Beyond the plateau stage, the lightcurve is dominated by $^{56}$Co decay, which we plot for a range of $^{56}$Ni masses using eq. (\ref{eq:lni}) as indicated ({\it dotted lines}).}
\label{fig:popov}
\epsscale{1.0}
\end{figure}

To better demonstrate the impact of this energy injection on the plateau phase, we plot example lightcurves using the analytic results of \citet{pop93} in Figure \ref{fig:popov}. Beyond the plateau stage, the lightcurve may reveal a power-law decline from $^{56}$Co decay, which we include for a range of $^{56}$Ni masses, using equation (\ref{eq:lni}). We do not plot the effective temperature since it is nearly constant at $\sim6000\ {\rm K}$ throughout the plateau phase.


\section{Conclusions and discussion}
\label{sec:conclusions}

We presented a study of the effect of supernova fallback accretion onto newly born magnetars. The combination of spin, magnetic field, and fallback rate was used to calculate the time evolution of the magnetar spin, estimate how much material was accreted, and determine whether the magnetar can expel enough material via the propeller mechanism to prevent collapse to a black hole. Strong magnetic fields and short spin periods are generally more advantageous for hindering black hole formation (but as Figures \ref{fig:contour} and \ref{fig:contour2} show, there are subtle changes to this picture depending on details of the time-dependent accretion rate). Even in cases that avoid becoming black holes, \mbox{$\sim0.3M_\odot$} or more of supernova fallback material is accreted, so we expect magnetars formed in collapsing massive stars to be more massive than the canonical $\sim1.4M_\odot$ neutron star mass. As discussed in \S \ref{sec:introduction}, there are at least three observed cases in our Galaxy of magnetars associated with $\sim30-40M_\odot$ progenitors. The propeller mechanism suggests a natural connection for why neutron stars associated with massive progenitors should have magnetar-strength fields.

Quickly spinning magnetars have been discussed as promising candidate systems for gravitational wave production via the time-changing quadrupole moment created by dynamical or secular instabilities (see Corsi \& M\'{e}sz\'{a}ros 2009, Ott 2009, and references therein). We conclude that there are two main cases that may lead to the emission of gravitational waves when fallback accretion is important. In the first case, if the propeller mechanism is active (typically $B\gtrsim5\times10^{14}\ {\rm G}$), the magnetar must begin with a sufficiently short spin period by the process of cooling and contraction, as is found for some models explored by Ott et al. (2006, in particular, see their summary of $\beta$-values in Table 4). It will then emit gravitational waves until it is spun down by accretion on a timescale of $\sim10-100\ {\rm s}$.  Since low accretion rates would extend the timescale for gravitational wave emission, and these correspond to more energetic explosions, our model predicts that gravitational waves (if present) are most likely in the most energetic events that do not collapse to black holes. We note that in such cases a gamma-ray burst may be created directly by the magnetar, as explored by Metzger et al. (2010, and references therein). In the second case, when accretion occurs directly onto the magnetar surface, the magnetar is spun up sufficiently to emit gravitational waves. But as we discussed at the end of \S 5, this only proceeds until the magnetar collapses to a black hole after $\lesssim50-200\ {\rm s}$  (see eq. [\ref{eq:tgw}]). The formation of a black hole and its subsequent accretion may then power a gamma-ray burst, again predicting a possible correlation between gravitational wave emission and a powerful electromagnetic event. Note however that in this case the gravitational waves would precede any sort of launching of a relativistic jet (in contrast to the model of Piro \& Pfahl 2007, which predicts gravitational waves coincident with the prompt gamma-ray emission, although both processes can occur in the same event).

When a magnetar is in the propeller regime, the expelled material collides with supernova ejecta, shock-heating it and energizing the supernova. \citet{mae07} proposed that some ultraluminous supernovae may be explained by dipole emission from a rapidly spinning magnetar, which was worked out in detail by Kasen \& Bildsten (2010) and Woosley (2010). We emphasize that our model is very different from theirs. In their case the magnetar directly powers the observed supernova luminosity. In our case the spin energy is injected earlier, creating a faster evolving supernova. We explored two regimes where this energy input may have a direct observational consequence: (1) in the case of a low-mass ($\lesssim5M_\odot$), hydrogen-deficient envelope, the additional energy gives rise to a broadlined Type Ib/c supernova, or potentially, a hypernova, depending on the amount of $^{56}$Ni synthesized, and (2) in the case of a massive ($\gtrsim10M_\odot$), hydrogen-rich envelope, we predict an event similar to a Type IIP supernova, although brighter and with higher velocities.

Our predictions for the lightcurves of energetic supernovae are independent of the actual mechanism for injecting the energy, requiring only deposition at early times ($\ll t_d$). Therefore, independent of our specific model for how the energy is produced, Type IIP supernovae that have similar lightcurves as we demonstrate in \S 5.3, along with high velocities, indicate an exceptional amount of energy injection. Indeed, SN 2009kf has a luminosity that implies an explosion energy of $\approx2\times10^{52}\ {\rm ergs}$ \citep{bot10}. The databases of high-cadence transient surveys, such as the Palomar Transient Factory \citep{law09} or Pan-STARRS \citep{kai02}, may reveal a larger population of Type IIP supernovae that, although not as extreme as SN 2009kj, may still require energy input beyond what is typically available for a supernova.

\acknowledgments
We thank Lars Bildsten, Luc Dessart, Brian Metzger, Robert Quimby, Uli Sperhake, and Todd Thompson for their helpful suggestions. We also thank Evan O'Connor for providing core collapse models for initial fallback calculations and estimates of the importance of an accretion disk. This work was supported through NSF grants AST-0855535, PHY-0960291, and OCI-0905046, and by the Sherman Fairchild Foundation. A.L.P. was supported in part by NASA ATP grant NNX07AH06G.

\appendix

\section{Neutrino-Cooled Accretion Columns}

In cases where $r_c>r_m>R$, material is magnetically channeled before reaching the magnetar's surface. This is traditionally called an accretion column in the study of accreting, magnetized white dwarfs and neutron stars \citep{fra92}. For a dipole field, $\sin^2\theta/r$ is constant, so that the path of the flow is described by the equation $1/r_m=\sin^2\theta/r$ (for simplicity we assume an aligned rotator). At a radius $r$ from the magnetar, the material is squeezed into an area
\be
	A(r)\approx \pi r^2\sin^2\theta \approx \pi r^2(r/r_m).
	\label{eq:area}
\ee
Assuming that the flow comes in at approximately free-fall, the velocity and density are
\be
	v_{\rm in}=\lp \frac{2GM}{r}\rp^{1/2},
	\qquad
	\rho_0 = \frac{\dot{M}}{2Av_{\rm in}}.
	\label{eq:vandrho}
\ee
where the factor of two is because there are two poles. From this we estimate
\be
	v_{\rm in}= 1.8\times10^{10}M_{1.4}^{1/2}r_{12}^{-1/2}{\rm cm\ s^{-1}},
\ee
where $r_{12}=r/12\ {\rm km}$. Combining equations (\ref{eq:rm}), (\ref{eq:area}), and (\ref{eq:vandrho}), the density is
\be
	\rho_0 = 1.6\times10^{8}\mu_{33}^{4/7}M_{1.4}^{6/7}r_{12}^{-5/2}\dot{M}_{-2}^{5/7}\ {\rm g\ cm^{-3}}.
\ee
The flow will go through a shock before reaching the stellar surface. This is checked by estimating the Mach number $\mathcal{M}$ of the flow,
\be
	\mathcal{M}^2 = \frac{v_{\rm in}^2}{c_s^2} = \frac{2\cdot 2^{1/4}}{\gamma}\lp \frac{p_B}{p}\rp^{1/2}\lp \frac{r}{r_m}\rp^{5/4},
\ee
where $c_s$ is the speed of sound, $\gamma$ is the adiabatic coefficient, and $p_B/p$ is the ratio of the magnetic pressure to the pressure of the gas (including ideal gas, radiation, and degeneracy contributions). For typical parameters,
\be
	\mathcal{M}
	= 1.4\mu_{33}^{5/7}M_{1.4}^{4/28}\dot{M}_{-2}^{5/14}r_{12}^{5/4}\lp \frac{p_B}{p}\rp^{1/2},
\ee
where we set $\gamma=4/3$ for a radiation dominated gas. In comparison, \citet{zd10} find that $\mathcal{M}\ll1$ because they assume much larger densities of $\sim10^{12}\ {\rm g\ cm^{-3}}$, but from continuity this implies an infall velocity much less than freefall ($v_{\rm in}\sim10^5\ {\rm cm\ s^{-1}}$) contrary to our expectations. In addition, we estimate the mean free path for proton-proton collisions to be $\sim10^{-7}\ {\rm cm}$ (using eq. [3.20] from Frank et al. 1992), much less than the width of the accretion column, so we expect the shock to be collisional.

The flow can be broken into two main regions. The first is a supersonic flow starting from the edge of the magnetosphere and then moving toward the magnetar pole. Then there is a shock, below which the subsonic flow settles onto the star. The jump conditions at the shock interface are
\be
	\rho_{\rm sh} = 7\rho_0, \qquad p_{\rm sh}=\frac{6}{7}\rho_{\rm sh}v_{\rm in}^2, \qquad v_{\rm sh}=\frac{1}{7}v_{\rm in},
\ee
for a strong shock with $\gamma=4/3$.
Therefore the post-shock density is
\be
	\rho_{\rm sh} = 1.1\times10^9\mu_{33}^{4/7}M_{1.4}^{6/7}r_{12}^{-5/2}\dot{M}_{-2}^{5/7}\ {\rm g\ cm^{-3}},
\ee
and
\be
	T_{\rm sh} &= &\lp \frac{3p_{\rm sh}}{a}\rp^{1/4}
	\nonumber
	\\
	 &=& 1.1\times10^{11}\mu_{33}^{1/7}M_{1.4}^{13/28}r_{12}^{-7/8}\dot{M}_{-2}^{5/28}\ {\rm K},
\ee
is the post-shock temperature.

The radiative diffusion timescale is approximately $t\sim \kappa\rho r/c\sim 10^4\ {\rm s}$, so the flow cannot cool via photons. Instead we consider neutrino cooling via electron-positron pair annihilation, which is given by \citep{pwf99}
\be
	\dot{q}_{\rm pairs} = 5\times10^{33}T_{11}^9\ {\rm ergs\ cm^{-3}\ s^{-1}}, 
\ee
where $T_{11}=T/10^{11}\ {\rm K}$. The timescale for this cooling is
\be
	t_{\rm pairs} &=& \frac{aT^4}{\dot{q}_{\rm pairs}}
	\nonumber
	\\
	&=& 5.2\times10^{-4}\mu_{33}^{-5/7}M_{1.4}^{-65/28}r_{12}^{35/8}\dot{M}_{-2}^{-25/28}\ {\rm s}.
\ee
It is also possible that Urca cooling is important, given by
\be
	\dot{q}_{\rm Urca} = 9\times10^{33}\rho_{10}T_{11}^6\ {\rm ergs\ cm^{-3}\ s^{-1}},
\ee
for a composition  of protons and neutrons (at these high temperature helium is photodisintegrated). The timescale for this cooling is
\be
	t_{\rm Urca} &=& \frac{aT^4}{\dot{q}_{\rm Urca}}
	\nonumber
	\\
	&=& 6.3\times10^{-4}\mu_{33}^{-6/7}M_{1.4}^{-28/14}r_{12}^{17/4}\dot{M}_{-2}^{-15/14}\ {\rm s}.
\ee
Urca cooling dominates when $t_{\rm Urca}<t_{\rm pair}$, implying an accretion rate
\be
	\dot{M} > 2.9\times10^{-2}\mu_{33}^{-4/5}M_{1.4}^{9/5}r_{12}^{7/10}\ M_\odot\ {\rm s^{-1}}.
\ee
This is a rather high accretion rate in comparison to what we consider, so it is sufficient to focus on pair cooling. The height of the shock above the magnetar surface is
\be
	H_{\rm sh} \approx v_{\rm sh} t_{\rm pair}
	= 1.3\mu_{33}^{-5/7}M_{1.4}^{-51/28}r_{12}^{31/8}\dot{M}_{-2}^{-25/28}\ {\rm km}.
	\nonumber
	\\
	\label{eq:hsh}
\ee
The shock therefore occurs at a radius of $r_{\rm sh}=R+H_{\rm sh}$. Note that the $r$ on the righthand side of equation (\ref{eq:hsh}) corresponds to $r_{\rm sh}$, so this equation is only accurate as long as $H_{\rm sh}\lesssim r_{\rm sh}\approx R$.
\begin{figure}
\epsscale{0.6}
\plotone{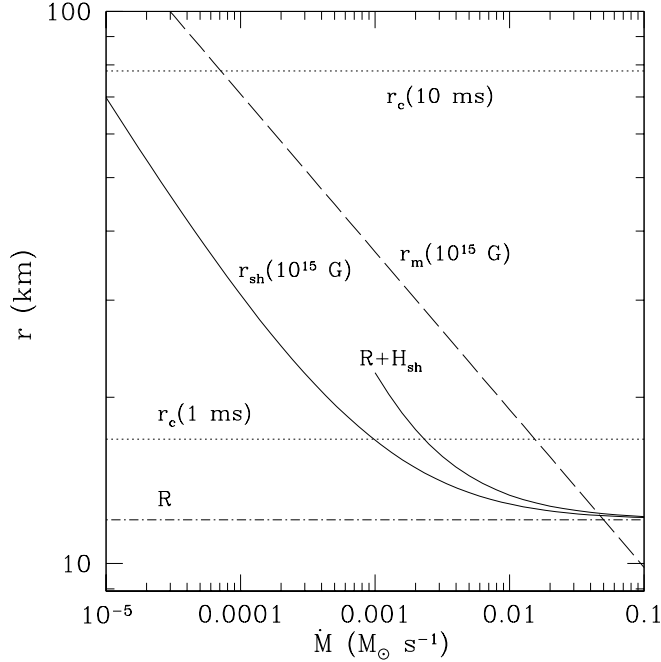}
\caption{Critical radii as a function of accretion rate for the problem of a neutrino-cooled accretion column. A given fluid element moves inward in radius (from top to bottom on the plot) at fixed accretion rate. In this way, one can read off what processes the fluid element experiences. When it reaches $r_m$ ({\it dashed line}), its motion is determined by the magnetic field. If $r_m>r_c$, then it will be expelled (we plot $r_c$ for 1 and 10 ms spin periods as examples; {\it dotted lines}). If $r_m<r_c$, then the flow will be channeled toward the magnetar pole and undergo a shock at $r_{\rm sh}$ ({\it solid line}). It finally reaches the magnetar surface at $R$ ({\it dot-dashed line}). As a comparison we also plot the approximation give by eq. (\ref{eq:hsh}) as the line labeled $R+H_{\rm sh}$.}
\label{fig:funnel}
\epsscale{1.0}
\end{figure}

When $H_{\rm sh}\gtrsim R$, we need to take into account adiabatic compression of material as it moves toward the magnetar pole. This gives a higher temperature at the magnetar surface in comparison to the shock radius, by an amount $T=T_{\rm sh}(R/\rsh)^{-1}$, which revises the pair cooling timescale by a factor of $(R/\rsh)^5$. We again write an equation for the shock height,
\be
	H_{\rm sh}= r_{\rm sh}-R \approx v_{\rm sh}t_{\rm pair}(r_{\rm sh}).
\ee
This expression can be solved numerically for $r_{\rm sh}$, which we plot in Figure \ref{fig:funnel} in comparison to other critical radii. This shows that for all accretion rates of interest, $r_{\rm sh}<r_m$, so the location of the shock is always within where funneled flow begins, as needed for consistency.


\end{document}